\PassOptionsToPackage{prologue,table}{xcolor}
\documentclass[10pt,sigconf,letterpaper,nonacm]{acmart}

\usepackage{colortbl}
\usepackage{tabularx}

\usepackage{booktabs}
\usepackage{multirow}
\usepackage[compatibility=false]{caption}
\usepackage{times}
\usepackage{url}
\usepackage[utf8]{inputenc}
\usepackage{graphicx}
\usepackage{amsmath}
\usepackage{amsfonts} 
\usepackage{tcolorbox}

\usepackage{siunitx}
\usepackage{adjustbox}

\AtBeginDocument{%
  }

\setcopyright{acmlicensed}
\copyrightyear{2025}
\acmYear{2025}
\acmDOI{}
\acmConference[IMC '25]{the 2025
ACM Internet Measurement Conference}{October 28--31,
  2025}{Madison, WI}
\acmISBN{}

\begin{document}

\title{REAL-IoT: Characterizing GNN Intrusion Detection
Robustness under Practical Adversarial Attack}

\author{Zhonghao Zhan}
\affiliation{%
  \institution{Imperial College London}
  \city{London}
  \country{UK}
  }

\author{Huichi Zhou}
\affiliation{%
  \institution{Imperial College London}
  \city{London}
  \country{UK}
  }

\author{Hamed Haddadi}
\affiliation{%
  \institution{Imperial College London}
  \city{London}
  \country{UK}
  }

\renewcommand{\shortauthors}{}

\begin{abstract}
Graph Neural Network (GNN)-based network intrusion detection systems (NIDS) are often evaluated on single datasets, limiting their ability to generalize under distribution drift. Furthermore, their adversarial robustness is typically assessed using synthetic perturbations that lack realism. This measurement gap leads to an overestimation of GNN-based NIDS resilience.
To address the limitations, we propose \textbf{REAL-IoT}, a comprehensive framework for robustness evaluation of GNN-based NIDS in IoT environments. Our framework presents a methodology that creates a unified dataset from canonical datasets to assess generalization under drift. In addition, it features a novel intrusion dataset collected from a physical IoT testbed, which captures network traffic and attack scenarios under real-world settings. Furthermore, using REAL-IoT, we explore the usage of Large Language Models (LLMs) to analyze network data and mitigate the impact of adversarial examples by filtering suspicious flows.
Our evaluations using REAL-IoT reveal performance drops in GNN models compared to results from standard benchmarks, quantifying their susceptibility to drift and realistic attacks. We also demonstrate the potential of LLM-based filtering to enhance robustness. These findings emphasize the necessity of realistic threat modeling and rigorous measurement practices for developing resilient IoT intrusion detection systems.
\end{abstract}



\maketitle

\section{Introduction}

The Internet of Things (IoT) represents a shift in connectivity, introducing a vast network of interconnected devices into modern life, spanning from smart homes and healthcare to industrial automation~\cite{IoTSurveyRG25}. Projections indicate billions of such devices communicating and sharing data, generating considerable volumes of network traffic~\cite{IoTSecuritySciSpace25}. However, this expansion simultaneously broadens the attack surface, introducing significant cybersecurity challenges~\cite{IoTSurveyRG25}. IoT devices are often resource-constrained (e.g., limited power, processing, and memory), highly heterogeneous in hardware and protocols, and frequently lack fundamental security measures from their design and manufacturing stages~\cite{IoTSecuritySciSpace25}. These inherent weaknesses make them targets for adversaries, posing substantial risks to user privacy, data integrity, and even physical safety through the manipulation of cyber-physical systems~\cite{IoTSurveyRG25}.

To counter these threats, Graph Neural Networks (GNNs) have emerged as a promising direction to develop advanced Network Intrusion Detection Systems (NIDS) tailored for IoT environments~\cite{Jahin25}. GNNs possess the unique ability to utilize network topology to model complex relational dependencies and structural patterns within network traffic data, potentially identifying sophisticated attacks that traditional methods might miss~\cite{GNNSurveyRG23}. However, the practical deployment of GNN-based NIDS faces significant hurdles. REAL-IoT is a comprehensive framework designed to systematically characterize the shortcomings of current GNN-based NIDS. 

\textbf{Susceptibility to Distribution Drift.} A primary challenge arises from the common practice of training and validating these models under the assumption of a static environment, often using isolated, single datasets~\cite{PierazziNDSS25}. This approach fails to account for the dynamic nature of real-world networks, leading to a phenomenon known as distribution drift or concept drift. Network traffic patterns evolve due to factors such as updates of devices, changes in user behavior, as well as the emergence of new attack techniques~\cite{PierazziNDSS25}. Consequently, GNN models trained on historical data often show poor generalization and performance degradation when confronted with these real-world distribution shifts, as demonstrated by studies showing substantial performance drops due to temporal and spatial transferability problems~\cite{PierazziNDSS25}. A model susceptible to such natural variations is unlikely to be reliable in practice.

\textbf{Unrealistic Adversarial Robustness Evaluation.} A second critical problem lies in the current methodologies used to evaluate the adversarial robustness of GNN-based NIDS~\cite{GNNSurveyRG23}. Robustness assessments are predominantly based on the generation of synthetic adversarial perturbations using white-box or black-box techniques~\cite{PracticalAttacksNeurIPS20}. White-box attacks, assuming that the adversary possesses complete knowledge of the target model's architecture and parameters, offer insights into worst-case vulnerabilities, but represent an unrealistic scenario for most external attackers~\cite{PracticalAttacksNeurIPS20}. Black-box attacks, while operating under more realistic knowledge constraints, often generate perturbations that lack practical feasibility within the network domain. These synthetic modifications may violate the inherent limitations of network protocols, ignore feature dependencies, or produce traffic patterns easily detectable by simple statistical analysis or other orthogonal security mechanisms~\cite{PracticalAttacksNeurIPS20}. Furthermore, GNNs introduce unique vulnerabilities related to the graph structure itself~\cite{GNNSurveyRG23}. Manipulation of the graph structure might be a more feasible and stealthy attack vector in dynamic IoT environments compared to constrained feature manipulation, yet such practical structural attacks are underexplored~\cite{GNNSurveyRG23}. There exists a critical gap in the measurement of the robustness of GNNs against practical adversarial attacks, particularly those tailored to the unique vulnerabilities of IoT environments~\cite{CelikNDSS21}.

\textbf{Potential Risks.} The confluence of these two problems, fragility to distribution drift and reliance on unrealistic adversarial evaluations, creates a significant risk. Models vulnerable to natural data shifts are likely even more susceptible to subtle attacks that mimic these shifts. Current evaluation practices often assess drift and adversarial robustness in isolation, overlooking the compounded risk and the potential for attacks to exploit natural data shifts. Therefore, the current measurement gap leads to an overestimation of the resilience of GNN-based NIDS, promoting a false sense of security regarding their deployment in IoT.

\textbf{Proposed Solution.} To bridge this critical measurement gap, we introduce REAL-IoT, a novel framework meticulously designed to facilitate a more realistic and holistic evaluation of the robustness of GNN-based IDS specifically for IoT environments. REAL-IoT addresses the limitations of current practices through the following contributions:
\begin{enumerate}
    \item \textbf{Unified Canonical Dataset Methodology:} Development of a methodology for standardizing canonical intrusion datasets, enabling GNN model evaluation both individually and on a merged, unified benchmarking dataset. This dataset is carefully designed to ensure a balanced representation of diverse attack types. It presents a rigorous assessment of model generalization and robustness.
    \item \textbf{Comprehensive Evaluation Methodology:} Definition of an integrated protocol for comprehensively evaluating GNN models. This includes assessing baseline performance, generalization under the aforementioned drift scenarios, and robustness against a variety of synthetic adversarial attacks (both white-box and black-box). Our empirical assessments quantify the performance discrepancies of current models compared to traditional evaluations, validating the need for such enhanced benchmarking.
    \item \textbf{Novel Dataset generated by physical IoT Testbed:} Construction of a physical IoT testbed and collect a novel network intrusion dataset from real-world security experiments. This dataset specifically targets practical adversarial attacks, enabling the measurement of GNN model performance under real-world conditions, moving beyond simulation-based assumptions. We also document the end-to-end testbed construction to ensure reproducibility.
    \item \textbf{LLM-Based Mitigation Analysis:} Introduction and evaluation of an approach leveraging LLMs as simulated cybersecurity analysts. These LLMs review network flow data to identify and flag suspicious or adversarially perturbed traffic, serving as a preemptive sanitization filter to enhance the robustness and generalization of downstream GNN detectors.

\end{enumerate}

The rest of the paper is structured as follows. It proceeds with background (Section 2), motivations (Section 3), the REAL-IoT framework and results (Section 4), and the controlled physical IoT testbed (Section 5). Section 6 concludes the paper. Anonymized artifacts are available. Appendices provide details for reproducibility.

\begin{table*}[!ht]
  \centering

  \caption{Benchmark Survey of GNN Models for Internet Intrusion Detection}
  \rowcolors{2}{gray!10}{white} 
  \begin{adjustbox}{max width=\linewidth}
  \begin{tabularx}{\textwidth}{
    >{\raggedright\arraybackslash}p{3cm}
    >{\centering\arraybackslash}p{1.6cm}
    >{\centering\arraybackslash}p{1.4cm}
    >{\centering\arraybackslash}X
    >{\centering\arraybackslash}p{1.4cm}
    >{\centering\arraybackslash}p{1cm}
    >{\centering\arraybackslash}p{0.6cm}
    >{\centering\arraybackslash}p{1.7cm}
  }
    \toprule
    \rowcolor{white}
    \textbf{Model} 
      & \textbf{Embedding} 
      & \textbf{Data Type} 
      & \textbf{Datasets} 
      & \textbf{Learning} 
      & \textbf{Open Source} 
      & \textbf{Year} 
      & \textbf{Comments} \\
    \midrule
    E-ResGAT~\cite{Chang21} 
      & Node, Edge & Std, NetFlow 
      & UNSW-NB15, CIC-DarkNet, CSE-CIC-IDS, ToN-IoT 
      & Supervised & Yes & 2021 
      & Code outdated \\

    E-GraphSAGE~\cite{Lo22} 
      & Node, Edge & Std, NetFlow 
      & ToN-IoT, NF-TON-IoT, BoT-IoT, NF-BoT-IoT 
      & Supervised & Yes & 2022 
      & Selected \\

    GLD-Net~\cite{Guo22} 
      & Node, Edge & Standard 
      & NSL-KDD2009, CIC-IDS2017 
      & Supervised & No & 2022 
      & Not open-sourced \\

    FT-GCN~\cite{Li22} 
      & Node, Edge & Std, NetFlow 
      & UNSW-NB15, CIC-Darknet2020, ISCXTor2016 
      & Supervised & No & 2022 
      & Not open-sourced \\

    GNN-NIDS~\cite{PujolPerich23} 
      & Node, Edge & Standard 
      & CIC-IDS2017 
      & Supervised & Yes & 2023 
      & Code outdated \\

    Anomal-E~\cite{Caville23} 
      & Node, Edge & NetFlow 
      & NF-UNSW-NB15-v2, NF-CSE-CIC-IDS2018-v2 
      & Self-supervised & Yes & 2023 
      & Selected \\

    PPT-GNN~\cite{VanLangendonck24} 
      & Node       & NetFlow 
      & NF-UNSW-NB15, NF-ToN-IoT, NF-BoT-IoT 
      & Self-supervised & Yes & 2024 
      & Tailored preprocess \\

    GNN-IDS~\cite{Sun24} 
      & Node       & Standard 
      & Synthetic, CICIDS-2017 
      & Supervised & Yes & 2024 
      & Partial code only \\

    XG-NID~\cite{Farrukh24} 
      & Node, Edge & Std, NetFlow 
      & CIC-IoT2023 
      & Supervised & Yes & 2024 
      & Partial code only \\

    FN-GNN~\cite{TranPark24} 
      & Node, Edge & Standard 
      & CICIDS-2017, UNSW-NB15 
      & Supervised & No & 2024 
      & Not open-sourced \\

    CAGN-GAT~\cite{Jahin25} 
      & Node       & Standard 
      & NSL-KDD, UNSW-NW15, CICIDS2017, KDD99 
      & Supervised & Yes & 2025 
      & Selected \\
    \bottomrule
  \end{tabularx}
  \end{adjustbox}
  \label{tab:gnn_benchmark}
\end{table*}

\section{Background and Challenges}
\subsection{GNNs for Network Intrusion Detection}
\label{subsubsec:gnns_for_nids}

The NIDS research landscape features a diverse collection of GNN architectures~\cite{GNNSurveyRG23}. Table~\ref{tab:gnn_benchmark} provides a comparative overview of several representative GNN-based NIDS models, highlighting this diversity. During the selection process, we surveyed models from peer-reviewed publications and conference proceedings that are publicly available. All listed models were carefully reviewed, and replication was attempted. We could not include most models in this study due to inaccessible code sources, lack of maintenance or updates, or the use of tailored preprocessing on benchmarking datasets, which could create an unfair comparison with other GNN-based NIDS models. After the survey and pre-screening, we selected three publicly available models for our in-depth evaluation within the REAL-IoT framework. These models are distinguished by their diverse methodologies:

\begin{itemize}
    \item \textbf{E-GraphSAGE}~\cite{Lo22} incorporates edge features to better model dynamic interactions. Selected for its focus on edge features in NIDS.
    \item \textbf{Anomal-E}~\cite{Caville23} uses self-supervised learning to create edge embeddings for anomaly detection without labels. Chosen for its suitability in data-scarce settings.
    \item \textbf{CAGN-GAT Fusion}~\cite{Jahin25} integrates Graph Attention Networks with contrastive learning for robust topology modeling. Selected to test its hybrid design for NIDS resilience.
\end{itemize}

\subsection{Canonical Datasets in NIDS Evaluation}
\label{subsubsec:common_datasets_nids} 

The evaluation of GNN-based NIDS typically relies on a range of established benchmark datasets.
However, as detailed in Table~\ref{tab:canonical_datasets}, these widely-used datasets collectively suffer from significant limitations that can compromise the validity and generalizability of evaluation results.
Common issues include outdated traffic patterns and attack vectors, insufficient diversity, unrealistic attack simulations, problematic labeling, and inconsistencies in feature generation across different sources~\cite{CICIDS17, AdversarialChallengesArXiv24}.
Such widespread shortcomings emphasize the critical need for standardized and more representative dataset construction methodologies, a core motivation for the unified dataset approach within our REAL-IoT framework.

\begin{table*}[!ht]
  \centering

  \caption{Overview of Canonical Intrusion Detection Datasets}
  \rowcolors{2}{gray!10}{white}
  \begin{adjustbox}{max width=\linewidth}
    \begin{tabularx}{\textwidth}{
      >{\raggedright\arraybackslash}p{3cm}
      >{\centering\arraybackslash}p{1.2cm}
      >{\centering\arraybackslash}p{2cm}
      >{\raggedright\arraybackslash}X
      >{\centering\arraybackslash}p{1.8cm}
      >{\raggedright\arraybackslash}X
    }
      \toprule
      \rowcolor{white}
      \textbf{Dataset Name}
        & \textbf{Year}
        & \textbf{Environment}
        & \textbf{Key Attack Types}
        & \textbf{Format}
        & \textbf{Known Limitations} \\
      \midrule
      KDD Cup 1999~\cite{KDD99}
        & 1999
        & General Network
        & DoS, Probe, U2R, R2L
        & Flow (CSV)
        & Outdated traffic and attack types; redundant records; unrealistic traffic mix~\cite{CICIDS17} \\

      NSL-KDD~\cite{NSLKDD09}
        & 2009
        & General Network
        & DoS, Probe, U2R, R2L
        & Flow (CSV)
        & Improved de-duplication but still outdated; potential class imbalance~\cite{OptimizingIoTIntrusionPMC24} \\

      UNSW-NB15~\cite{UNSWNB15}
        & 2015
        & General Network
        & Fuzzers, Analysis, Backdoors, DoS, Exploits, Generic, Recon, Shellcode, Worms
        & Flow/PCAP
        & Lab-generated; possible feature inconsistencies; class imbalance~\cite{SurveyDataDrivenNIDS22} \\

      CIC-IDS2017~\cite{CICIDS17}
        & 2017
        & General Network
        & Brute Force (FTP/SSH), DoS, Heartbleed, Web Attacks, Infiltration, Botnet, DDoS
        & Flow/PCAP
        & Realistic background traffic but potential labeling and feature inconsistencies~\cite{CICIDS17} \\

      CIC-IDS2018~\cite{CICIDS18}
        & 2018
        & General Network
        & Brute Force, DoS, DDoS, Web Attacks, Infiltration, Botnet
        & Flow/PCAP
        & Extension of 2017 dataset with larger volume; similar limitations~\cite{UnifiedDeepLearningRG21} \\

      Bot-IoT~\cite{BotIoT19}
        & 2018
        & IoT (Smart Home)
        & DoS/DDoS (TCP/UDP/HTTP), Recon, Data/Theft (Keylogging)
        & Flow/PCAP
        & Realistic IoT testbed; focused scenarios; class imbalance~\cite{LaunchingAdversarialAttacksMDPI21} \\

      ToN\_IoT~\cite{ToNIoT20}
        & 2019
        & IoT/IIoT (Multiple)
        & DoS, DDoS, Scanning, Ransomware, Backdoor, Injection, MitM, XSS
        & Flow (CSV)
        & Potential cross-source inconsistencies~\cite{ExpectationsVsRealityArXiv24} \\
      \bottomrule
    \end{tabularx}
  \end{adjustbox}
  \label{tab:canonical_datasets}
\end{table*}

\subsection{Distribution Drift in GNN}
\label{subsubsec:distribution_drift} 

A critical challenge in deploying machine learning models for cybersecurity, including GNN-based NIDS, is distribution drift~\cite{METANOIAArXiv25}.
This phenomenon refers to changes in the statistical properties of data over time, rendering models trained on historical data less effective on current or future data.

The consequence of unaddressed distribution drift is significant performance degradation for NIDS, leading to increased false alarms and missed detections, as empirically demonstrated in studies using time-separated data~\cite{PierazziNDSS25}.
While GNNs are experts at modeling complex network structures, they are not inherently immune to such drift, which can affect input features, graph topology, and feature-label relationships~\cite{METANOIAArXiv25}.
Therefore, evaluating GNN-based NIDS performance under conditions that simulate distribution drift is vital for realistic robustness assessment as well. It is a key consideration in the design of our REAL-IoT framework and an area requiring further investigation in cybersecurity~\cite{DesigningReliableGFMArXiv25}.

\subsection{Adversarial Attacks against GNNs}
\label{subsubsec:adversarial_attacks_gnns} 

Adversarial machine learning investigates how models can be manipulated by malicious inputs, with evasion attacks being particularly relevant for NIDS evaluation~\cite{GraphRobustnessBookChapter, NISTAMLTaxonomy23}.
The feasibility and potential of these attacks heavily depend on the adversary's knowledge of the target model, commonly categorized ~\cite{PracticalAttacksNeurIPS20} as:
\begin{itemize}
    \item \textbf{White-box attacks}, assuming full knowledge (architecture, parameters, training data), enabling powerful gradient-based methods like Projected Gradient Descent (PGD)~\cite{Madry18_PGD}. While offering insights into worst-case vulnerabilities, this assumption is often unrealistic for external attackers.
    \item \textbf{Black-box attacks}, where the adversary only queries the model and observes outputs. Attacks often rely on query-based optimization, or score-based methods ~\cite{UNSWNB15}. These are more realistic but typically less effective or require extensive querying.
    \item \textbf{Gray-box attacks}, representing partial knowledge (e.g., model architecture but not parameters).
\end{itemize}

Applying these adversarial concepts to GNN-based NIDS, however, presents unique and significant challenges~\cite{PracticalAttacksNeurIPS20}.
Network data is inherently structured, discrete, and governed by strict protocols; naive feature perturbations often result in invalid or easily detectable traffic, diminishing the practical relevance of many synthetic attacks.
GNNs further introduce vulnerabilities rooted in their graph structures, where graph manipulations can directly undermine their core reasoning mechanisms~\cite{GNNSurveyRG23}.
Despite the stealth and effectiveness of such attacks, research into realistic and feasible structural threats against GNN-based NIDS within IoT environments remains insufficiently explored.
This critical gap in evaluating GNNs under practical adversarial settings undermines our trust in their robustness in real-world deployments.

\subsection{Benchmarking GNN Robustness}
\label{subsubsec:benchmarking_robustness} 

Robust and comprehensive benchmarking practices for GNN-based NIDS in dynamic IoT environments also require more attention~\cite{AdversarialChallengesArXiv24, AreWeThereYetArXiv25}.While general frameworks for adversarial machine learning and GNN robustness exist, they often focus on generic graph tasks or overlook the unique operational context of NIDS. Furthermore, standard evaluation metrics require careful adaptation for network and graph-structured data, where simple perturbations in other domains may not apply~\cite{GraphRobustnessBookChapter}.

Consequently, there is a critical need for specialized benchmarking methodologies that can holistically assess GNN-based NIDS performance under practical conditions relevant to the IoT landscape, a gap our REAL-IoT framework aims to address.

\begin{figure*}[t!]
  \centering
  \includegraphics[width=0.75\linewidth]{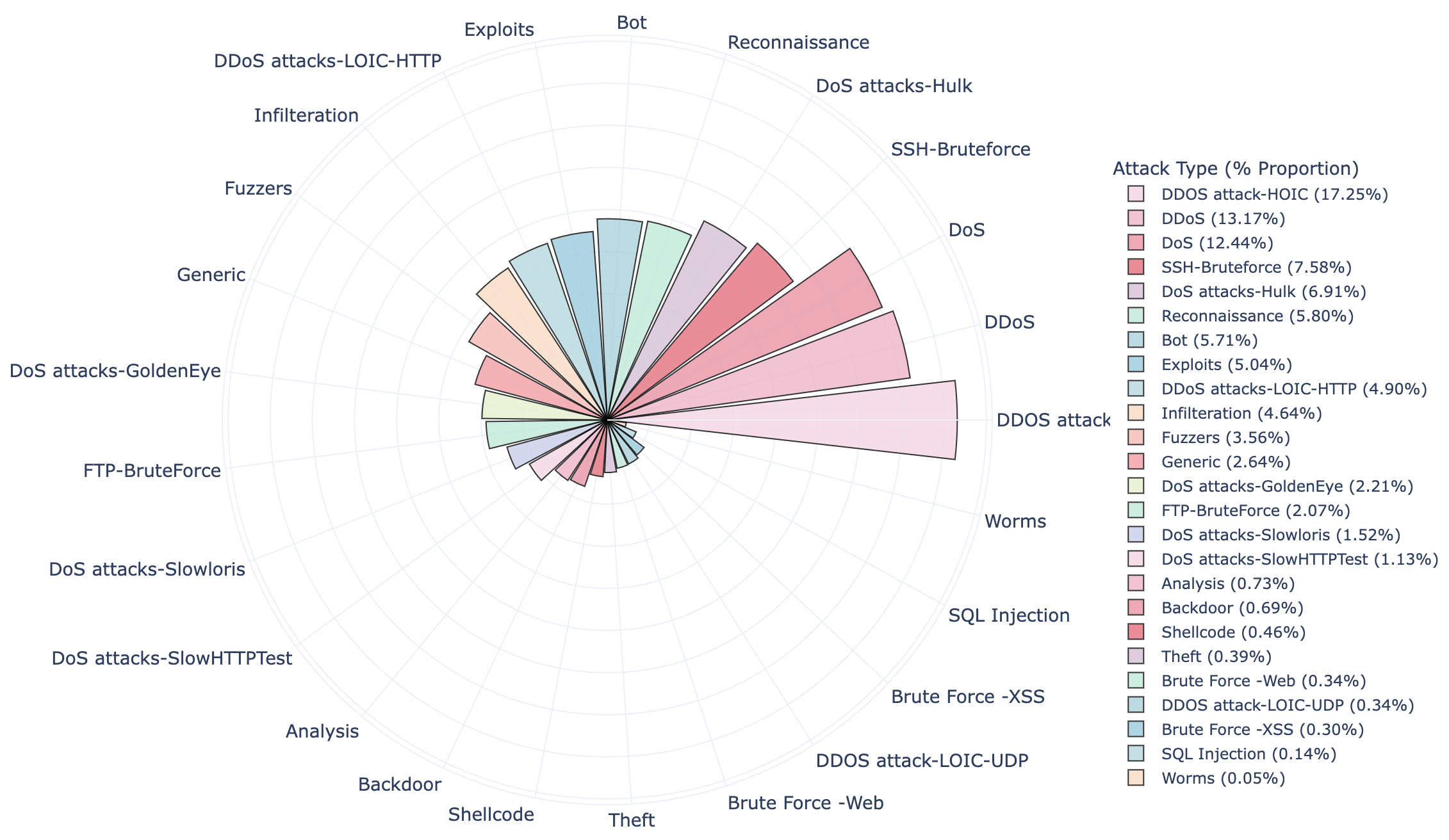}
  \label{fig: overview}
  \caption{The intrusion types of unified dataset for distribution drift test}
\end{figure*}

\section{Motivations}

The Introduction identified key challenges in evaluating GNN-based IDS, such as distribution drift and unrealistic adversarial assessments~\cite{PierazziNDSS25,PracticalAttacksNeurIPS20}. Section 2 detailed their roots in static datasets and generic evaluation methods. This section explains why these issues are particularly critical in IoT environments, motivating the need for the REAL-IoT framework to ensure robust GNN evaluations in real-world IoT deployments.

\textbf{Dynamic IoT Topologies and Resource Constraints.} Unlike static datasets such as CIC-IDS2017~\cite{CICIDS17}, real IoT environments exhibit unpredictable topology shifts that degrade GNN performance. IoT devices’ limited computational power, memory, and energy also restrict frequent model retraining or complex drift detection~\cite{IoTSecuritySciSpace25}. Our controlled physical testbed, using Raspberry Pis to simulate Smart Home device traffic, reveals GNNs’ vulnerability to such drift under resource constraints, emphasizing the need for IoT-specific evaluation methodologies.

\textbf{IoT-Specific Traffic Patterns.} IoT traffic, such as MQTT-based flows, is often bursty and event-driven. It differs significantly from general network datasets~\cite{AreWeThereYetArXiv25}. Standard benchmarks like UNSW-NB15~\cite{UNSWNB15} fail to capture these patterns. Adversarial assessments often rely on generic perturbations irrelevant to IoT's constrained protocols~\cite{PracticalAttacksNeurIPS20}. By building a controlled IoT physical testbed, we collect MQTT traffic data that enable evaluations in more realistic IoT behaviors, addressing this measurement gap.

\textbf{Holistic and Realistic Measurement.} The interplay of dynamic drift and generic evaluations overestimates GNN resilience in IoT settings. REAL-IoT bridges this gap with a unified dataset methodology and a physical testbed, ensuring GNNs are tested under conditions mirroring operational IoT networks. This framework provides reliable insights into model robustness for resilient IDS deployments.

\section{The REAL-IoT Framework}

The REAL-IoT framework is proposed as a comprehensive solution for evaluating the robustness of GNN-based IDS, focusing on addressing challenges related to dataset heterogeneity, class imbalance, and standardized evaluation. It includes two main components: a Unified Canonical Dataset Generator and an Integrated Evaluation Engine. This section details these components and the subsequent evaluation methodologies employed to assess GNN models.

\subsection{Overall Architecture}

The REAL-IoT framework takes as input raw intrusion detection datasets and the GNN-based IDS models intended for evaluation. Its core components work collaboratively:
\begin{itemize}
    \item \textbf{Unified Canonical Dataset Generator:} This component first standardizes selected canonical NetFlow datasets, enabling individual GNN model evaluation on diverse, distinct data sources. It subsequently processes these into a large-scale, unified dataset, applying adaptive sampling to manage its size and mitigate class imbalance, thereby enabling realistic simulations of distribution drift.
    
    \item \textbf{Integrated Measurement Engine:} This engine executes a defined evaluation protocol. It utilizes the datasets generated by the first component (both individual and unified) and applies various assessment techniques. It utilizes the application of synthetic adversarial attacks and the analysis of LLM-based mitigation strategies to perform a multi-dimensional robustness assessment of the input GNN models.
    
\end{itemize}
The final output is a comprehensive robustness report detailing the model's performance across baseline conditions, distribution drift scenarios, synthetic adversarial attacks, and the effectiveness of evaluated mitigation techniques.

\subsection{Unified Dataset Construction}
\label{subsec:unified_dataset_construction}

This component demonstrates our methodology for constructing a unified evaluation framework, addressing dataset heterogeneity and class imbalance through systematic NetFlow standardization and adaptive sampling, and defining a consistent graph construction process.

\subsubsection{Dataset Selection}
\label{subsubsec:dataset_selection}

We select NetFlow datasets as the basis for our study. NetFlow provides a compact and structured summary of network communications, capturing essential flow-level statistics (IP addresses, ports, protocols, byte/packet counts, duration, flags) that are well-suited for graph-based modeling of network interactions \cite{Hofstede14_FlowBasedReview}. Our selected datasets represent diverse network environments and include contemporary attack scenarios:
\begin{itemize}
    \item \textbf{NF-BoT-IoT (v1 \& v2):} Featuring traffic from IoT environments, including DoS, DDoS, and information theft attacks. Version 2 offers extended NetFlow features but presents scalability challenges due to its size (several GBs) \cite{Sarhan22_StandardFeatureSet}. 
    \item \textbf{NF-CSE-CIC-IDS2018 (v2):} Including benign traffic and diverse modern attacks (Brute Force, DDoS, Web Attacks, Infiltration) captured within a realistic enterprise network topology \cite{Sarhan22_StandardFeatureSet}.
    \item \textbf{NF-UNSW-NB15 (v2):} Containing synthesized modern attack activities (Fuzzers, Backdoors, Exploits, Worms) alongside benign traffic \cite{Sarhan22_StandardFeatureSet}. 
\end{itemize}
These datasets, denoted as \( \mathcal{D}_{raw} \), serve as the input to our standardization process. They are processed by a unified script to ensure that every individual dataset is transformed into a consistent format with the adoption of a common feature schema (as shown by Table~\ref{tab:netflow_features_adjusted} in the Appendices) before direct use by the GNN models.

\begin{table*}[t]
\centering
\caption[GNN Model Performance Comparison Across Datasets (Short title for LoT if needed)]{GNN Model Performance Comparison Across Datasets. Values marked with * are reported from the respective model's original paper; these are directly adopted for fairness. E-GraphSAGE only reported Accuracy and Precision in its binary classification results but not multiclass classification. Therefore binary classification result is selected. Anomal-E’s precision was not reported in its publication (indicated by – with a *). Given Anomal-E's multiple algorithm architecture, we pick the best performing algorithm (HBOS) and parameter for the data in this table.
}
\label{tab:gnn_performance_revised}

\sisetup{
  detect-weight        = true,
  detect-family        = true,
  round-mode           = places,
  round-precision      = 3,
  table-number-alignment = center
}
\begin{adjustbox}{width=\textwidth}
\rowcolors{2}{gray!10}{white}
\begin{tabular}{
  @{\kern\tabcolsep} l
  l
  S[table-format=1.3, table-align-text-post=false]
  S[table-format=1.3, table-align-text-post=false]
  S[table-format=1.3, table-align-text-post=false]
  <{\kern\tabcolsep}}
\toprule
\rowcolor{white}
\textbf{Model}
  & \textbf{Test Dataset Configuration}
  & {\textbf{Accuracy}}
  & {\textbf{Precision (Weighted)}}
  & {\textbf{F1-Score (Weighted)}} \\
\midrule
Anomal-E
  & Original Dataset (NF-UNSW-NB15-v2) & 0.987* & {--}* & 0.924* \\ 
  & Original Dataset (NF-CSE-CIC-IDS2018-v2) & 0.975* & {--}* & 0.881* \\
\cmidrule(lr){2-5}
  & Standardized NF-BoT-IoT v2 & 0.488 & 0.976 & 0.641 \\
  & Standardized NF-CICIDS2018 v2 & 0.945 & 0.977 & 0.977 \\
  & Standardized NF-UNSW-NB15 v2 (Full) & 0.974 & 0.972 & 0.973 \\
\cmidrule(lr){2-5}
  & Unified Dataset & 0.945 & 0.976 & 0.977 \\
\midrule
E-GraphSAGE
  & Original Dataset (NF-BoT-IoT) & 0.936* & 1.000* & 1.000* \\ 
  & Original Dataset (NF-ToN-IoT) & 0.997* & 1.000* & 1.000* \\
\cmidrule(lr){2-5}
  & Standardized NF-BoT-IoT v2 & 0.983 & 0.984 & 0.984 \\
  & Standardized NF-CICIDS2018 v2 & 0.640 & 0.968 & 0.736 \\
  & Standardized NF-UNSW-NB15 v2 & 0.974 & 0.987 & 0.980 \\
\cmidrule(lr){2-5}
  & Unified Dataset & 0.675 & 0.945 & 0.884 \\
\midrule
CAGN-GAT Fusion 
  & Original Dataset (KDD CUP 99) & 0.992* & 0.993* & 0.901* \\
  & Original Dataset (UNSW-NB15) & 0.995* & 0.962* & 0.918* \\
  & Original Dataset (CICIDS2017) & 0.985* & 0.984* & 0.946* \\
  & Original Dataset (NSL-KDD) & 0.987* & 0.980* & 0.984* \\
\cmidrule(lr){2-5}
  & Standardized NF-BoT-IoT v2 & 0.940 & 0.999 & 0.969 \\
  & Standardized NF-CICIDS2018 v2 & 0.929 & 0.854 & 0.908 \\
  & Standardized NF-UNSW-NB15 v2 & 0.696 & 0.102 & 0.182 \\ 
\cmidrule(lr){2-5}
  & Unified Dataset & 0.851 & 0.948 & 0.823 \\
\bottomrule
\end{tabular}
\end{adjustbox}
\end{table*}

\subsubsection{Adaptive Stratified Sampling}
\label{subsubsec:sampling}

Processing multi-gigabyte NetFlow datasets with GNNs can be computationally expensive. To create manageable yet representative subsets (\( \mathcal{D}_{\text{reduced}} \)), especially from large datasets like NF-BoT-IoT v2, we employ an adaptive stratified sampling strategy designed to mitigate class imbalance~\cite{Layeghy23_ClassImbalance}. This step is performed \emph{before} standardization if a reduced dataset is required to ensure the fidelity of sampling from the original datasets.

\begin{enumerate}
    \item \textbf{Class Distribution Analysis:} Compute the counts \( N_c \) for each class \( c \) (attack types and 'Normal') in the full dataset \( \mathcal{D}_{\text{full}} \), and the total number of records \( N_{\text{total}} = \sum_c N_c \).
    \item \textbf{Adaptive Sampling Rate Calculation:} Define class rarity thresholds based on relative frequency: \( \theta_{\text{rare}} = p_{\text{rare}} N_{\text{total}} \) and \( \theta_{\text{uncommon}} = p_{\text{uncommon}} N_{\text{total}} \). Given a target overall reduction rate \( r_{\text{base}} \) (e.g., 0.05 for 5\% retention), calculate the class-specific sampling rate \( r_c \) by adaptive multipliers \( M_{\text{rare}}, M_{\text{uncommon}}, M_{\text{common}} \) and a minimum retention rate \( r_{\text{high}} \):
    \begin{equation}
    \scriptsize
    \begin{aligned}
        r_c = \begin{cases} 
            \max(r_{\text{high}}, r_{\text{base}} \times M_{\text{rare}}) & \text{if } N_c < \theta_{\text{rare}} \\ 
            \min(1.0, r_{\text{base}} \times M_{\text{uncommon}}) & \text{if } \theta_{\text{rare}} \leq N_c < \theta_{\text{uncommon}} \\ 
            r_{\text{base}} \times M_{\text{common}} & \text{if } N_c \geq \theta_{\text{uncommon}} 
        \end{cases}
    \end{aligned}
    \end{equation}
    Additionally, if \( N_c \) is below an absolute threshold \( N_{\text{min}} \) (e.g., 1000), we set \( r_c = 1.0 \) to preserve extremely rare classes entirely. The multipliers are chosen heuristically based on observed class distributions to boost the retention of rare classes relative to common ones, aiming for a more balanced representation in \( \mathcal{D}_{\text{reduced}} \) while approximating the overall \( r_{\text{base}} \).
    \item \textbf{Stratified Sampling Implementation:} Process \( \mathcal{D}_{\text{full}} \) (potentially in chunks for scalability). For each class \( c \), randomly sample a fraction \( r_c \) of its instances. The union of these sampled instances forms the reduced dataset \( \mathcal{D}_{\text{reduced}} \). This ensures the expected number of instances for class \( c \) in the reduced set is \( \mathbb{E}[| \mathcal{D}_{\text{reduced}, c} |] = N_c \times r_c \).
\end{enumerate}
This stratified approach explicitly addresses class imbalance during reduction, ensuring that rare attack types are not disproportionately lost compared to uniform random sampling.

\subsubsection{NetFlow Standardization Pipeline}
\label{subsubsec:standardization}

The core of our methodology is a standardization pipeline designed to convert raw or reduced NetFlow datasets (\( \mathcal{D}_{\text{raw}} \) or \( \mathcal{D}_{\text{reduced}} \)) into a consistent tabular format \( \mathcal{D}'_{\text{std}} \). This involves several key steps, addressing challenges highlighted in prior work regarding dataset heterogeneity~\cite{CICIDS18, Viegas21_ReliableBenchmark}.

\begin{enumerate}

\item \textbf{Schema Mapping and Feature Harmonization:}
We define a unified target schema \( \mathcal{S} \) consisting of essential flow attributes, drawing inspiration from standardized sets like those proposed by Sarhan et al.~\cite{Sarhan22_StandardFeatureSet}. A mapping function \( \Phi_{\text{map}}: \mathcal{S}_{\text{raw}} \rightarrow \mathcal{S} \) translates columns from the raw dataset schema \( \mathcal{S}_{\text{raw}} \) to the target schema \( \mathcal{S} \). This includes:
\begin{itemize}
    \item Identifier Alignment: Mapping source/destination IPs and ports.
    \item Protocol Representation: Storing the numerical identifier and string name using a predefined mapping \( \mathcal{M}_{\text{proto}} = \{1:\text{``ICMP''}, 6:\text{``TCP''}, 17:\text{``UDP''}, \text{\dots}\} \).
    \item TCP Flag Handling: Using the standard TCP Flags field directly from NetFlow. For datasets derived from formats lacking this field, we reconstruct a comparable representation if possible.
\end{itemize}

\item \textbf{Label Unification:}
   Categorical Attack Type : We define a unified attack taxonomy \( \mathcal{T}_{\text{attack}} \) and map original attack category strings \( c_{\text{orig}} \) to standardized types \( c_{\text{std}} \in \mathcal{T}_{\text{attack}} \) using a carefully designed mapping \( \mathcal{M}_{\text{attack}}: \mathcal{C}_{\text{orig}} \rightarrow \mathcal{T}_{\text{attack}} \). This consolidates variations and handles dataset-specific categories. Unmappable types are assigned `Other'.

\item \textbf{Numerical Feature Scaling:}
To prevent features with larger ranges from dominating model training and to improve convergence, we apply Z-score standardization to the selected set of numerical features \( \mathbf{X}_{num} \subset \mathcal{D}'_{std} \). For each feature vector \( \mathbf{x} \in \mathbf{X}_{num} \), the standardized feature \( \mathbf{x}' \) is calculated as:
\begin{equation}
    \mathbf{x}' = \frac{\mathbf{x} - \boldsymbol{\mu}_{train}}{\boldsymbol{\sigma}_{train}}
\end{equation}
where \( \boldsymbol{\mu}_{train} \) and \( \boldsymbol{\sigma}_{train} \) are the element-wise mean and standard deviation vectors computed exclusively from the training portion of the dataset (\( \mathcal{D}'_{train} \)). These statistics are saved and consistently applied to the validation (\( \mathcal{D}'_{val} \)) and testing (\( \mathcal{D}'_{test} \)) sets to prevent data leakage. Z-score standardization is chosen for its robustness to outliers compared to Min-Max scaling and its common usage in deep learning pipelines \cite{Goodfellow16_DeepLearning}.

\end{enumerate}

\subsubsection{Standardized Graph Construction for Evaluation}
\label{subsubsec:graph_construction}

A critical component of our standardization pipeline is the consistent transformation of the processed tabular NetFlow data (\( \mathcal{D}'_{\text{std}} \) or \( \mathcal{D}'_{\text{reduced}} \)) into graph structures suitable for GNN ingestion. This step is critical for ensuring a fair and controlled measurement of the diverse GNN architectures under study. By defining a unified graph construction methodology, we provide all models with an equivalent structural representation of the network data, thereby isolating the GNN architecture itself as the primary variable of interest in our performance evaluation.

While individual GNNs might achieve different performance characteristics if paired with bespoke, model-specific graph construction heuristics \cite{Zhou20_GNNSurvey}, our comparative objective requires a common foundation. We adopt a widely recognized IP-centric communication graph model:

\begin{itemize}
    \item \textbf{Nodes \( \mathcal{V} \):} Each IP address observed in the standardized flow records constitutes a node \( v \in \mathcal{V} \). For the GNNs evaluated, node features were either initialized as learnable embeddings or simple uniform vectors as the primary discriminative information is encoded in edge attributes and graph structure.
    \item \textbf{Edges \( \mathcal{E} \):} Each standardized NetFlow record \( \mathbf{f} \in \mathcal{D}' \) is translated into a directed edge \( e_{uv} \in \mathcal{E} \), where the source node \( u \) corresponds to \( \mathbf{f}[\texttt{IPV4\_SRC\_ADDR}] \) and the destination node \( v \) to \( \mathbf{f}[\texttt{IPV4\_DST\_ADDR}] \). This directly models individual communication flows.
    \item \textbf{Edge Features \( \mathbf{x}_{e_{uv}} \):} The feature vector for an edge \( e_{uv} \) is directly derived from its corresponding standardized flow record \( \mathbf{f} \). This vector has the preprocessed numerical features (as outlined in Table~\ref{tab:netflow_features_adjusted}) that characterize the flow. The subset of standardized features utilized as edge attributes aligns with the input requirements of each GNN model, ensuring all models draw from the consistent feature set.
\end{itemize}
This standardized IP-centric graph construction provides a direct and interpretable representation of network interactions, which is fundamental to the integrity of REAL-IoT's measurement. It enables a robust and comparable assessment of GNN model capabilities on a fair ground, which fills a gap in contemporary GNN-based NIDS model measurements.

\subsection{Analysis of GNN Model Performance}

Table ~\ref{tab:gnn_performance_revised} presents an evaluation of selected GNN-based IDS models: \textbf{Anomal-E}, \textbf{E-GraphSAGE}, and \textbf{CAGN-GAT} across individual standardized datasets and a unified dataset. 

\textbf{Anomal-E} records an F1-Score of 0.641 on the standardized NF-BoT-IoT v2 dataset, which is significantly lower than its performance on other standardized and original datasets. This gap likely stems from the model’s unsupervised embedding technique, which may struggle to capture IoT-specific anomalous patterns. The NF-BoT-IoT v2 dataset poses a unique challenge due to its high volume of imbalanced IoT-related attacks, with a Benign to Attack ratio ranging from 0.03 to 9.97~\cite{Sarhan22_StandardFeatureSet}. However, on the unified dataset, Anomal-E achieves an impressive F1-Score of 0.977. This suggests that its hybrid approach effectively leverages diverse data to improve generalization. The improvement also highlights the effectiveness of our stratified sampling method in constructing the unified dataset, as it reduces the impact of class imbalance and boosts model robustness.

\textbf{E-GraphSAGE} exhibits limited adaptability. It achieves an accuracy of 0.640 on NF-CSE-CIC-IDS2018 v2, potentially due to its mean neighborhood aggregation method, which appears sensitive to differences in feature distributions and graph structures. On the unified dataset, its accuracy improves only marginally to 0.675, suggesting persistent challenges in handling the heterogeneous properties of combined data.

\textbf{CAGN-GAT Fusion} underperforms on NF-UNSW-NB15 v2, with an F1-Score of 0.182. This may result from the standardization process disrupting contextual features essential to its attention mechanism. While its performance on the unified dataset rises to an F1-Score of 0.823, it remains below Anomal-E, indicating partial resilience to data diversity but susceptibility to characteristic shifts.

These initial findings highlight the utility of the REAL-IoT's dataset generation in uncovering model-specific limitations and assessing robustness to distribution drift. To conduct a more exhaustive assessment, we now demonstrate our comprehensive evaluation methodology and experimental setup.

\begin{table}[ht]
  \centering
  \caption{Adversarial Attack Results of Anomal-E}
  \label{tab:adversarial_results_anomal}
  
  \setlength{\tabcolsep}{3.5pt}
  \renewcommand{\arraystretch}{1}
  
  \sisetup{
    detect-weight=true,
    detect-family=true,
    table-number-alignment = center,
    round-mode=places,
    round-precision=2,
  }
  \begin{tabular}{
    @{}
    l   
    l   
    S   
    S   
    S   
    S   
    S   
    @{}
  }
    \toprule
    \textbf{Attack Type} & \textbf{Param} 
      & {\textbf{Acc}} 
      & {\textbf{Prec}} 
      & {\textbf{Recall}} 
      & {\textbf{F1}} 
      & {\textbf{AUC}} \\
    \midrule
    Clean 
      & --       & 0.81 & 0.91 & 0.10 & 0.17 & 0.85 \\
    \midrule
    \multirow{4}{*}{PGD} 
      & $\epsilon=0.01$ & 0.81 & 0.91 & 0.10 & 0.17 & 0.74 \\
      & $\epsilon=0.05$ & 0.21 & 0.20 & 0.87 & 0.32 & 0.74 \\
      & $\epsilon=0.10$ & 0.21 & 0.21 & 1.00 & 0.35 & 0.73 \\
      & $\epsilon=0.20$ & 0.21 & 0.21 & 1.00 & 0.35 & 0.80 \\
    \midrule
    \multirow{4}{*}{EdgeRemove} 
      & 5\%   & 0.81 & 0.91 & 0.10 & 0.17 & 0.85 \\
      & 10\%  & 0.81 & 0.92 & 0.10 & 0.18 & 0.86 \\
      & 20\%  & 0.81 & 0.91 & 0.10 & 0.17 & 0.85 \\
      & 30\%  & 0.81 & 0.92 & 0.11 & 0.19 & 0.86 \\
    \midrule
    \multirow{3}{*}{NodeInject} 
      & 5\%   & 0.79 & 0.57 & 0.10 & 0.17 & 0.86 \\
      & 10\%  & 0.77 & 0.41 & 0.10 & 0.16 & 0.86 \\
      & 20\%  & 0.76 & 0.40 & 0.10 & 0.16 & 0.86 \\
    \bottomrule
  \end{tabular}
\end{table}

\begin{table}[ht]
  \centering
  \caption{Adversarial Attack Results of EGraphSage}
  \label{tab:adversarial_results_egraphsage}
  \setlength{\tabcolsep}{4pt}
  \renewcommand{\arraystretch}{1}
  \sisetup{
    detect-weight=true,
    detect-family=true,
    table-number-alignment = center,
    round-mode=places,
    round-precision=2,
  }
  \begin{tabular}{
    @{}
    l   
    l   
    S   
    S   
    S   
    S   
    S   
    @{}
  }
    \toprule
    \textbf{Attack Type} & \textbf{Param}
      & {\textbf{Acc}}
      & {\textbf{Prec}}
      & {\textbf{Recall}}
      & {\textbf{F1}}
      & {\textbf{AUC}} \\
    \midrule
    Clean
      & --           & 0.90 & 0.94 & 0.83 & 0.88 & 0.89 \\
    \midrule
    \multirow{3}{*}{PGD}
      & $\epsilon=0.3$ & 0.91 & 0.96 & 0.85 & 0.90 & 0.90 \\
      & $\epsilon=0.5$ & 0.88 & 0.90 & 0.85 & 0.87 & 0.90 \\
      & $\epsilon=0.7$ & 0.73 & 0.65 & 0.91 & 0.76 & 0.90 \\
    \midrule
    \multirow{4}{*}{EdgeRemove}
      & 5\%   & 0.90 & 0.94 & 0.83 & 0.88 & 0.89 \\
      & 10\%  & 0.90 & 0.94 & 0.83 & 0.88 & 0.89 \\
      & 20\%  & 0.90 & 0.94 & 0.83 & 0.89 & 0.89 \\
      & 30\%  & 0.90 & 0.94 & 0.83 & 0.89 & 0.89 \\
    \midrule
    \multirow{3}{*}{NodeInject}
      & 5\%   & 0.81 & 0.85 & 0.83 & 0.84 & 0.86 \\
      & 10\%  & 0.76 & 0.81 & 0.83 & 0.82 & 0.83 \\
      & 20\%  & 0.73 & 0.80 & 0.84 & 0.82 & 0.78 \\
    \bottomrule
  \end{tabular}
\end{table}

\begin{table}[ht]
  \centering
  \caption{Adversarial Attack Results of CAGN-GAT}
  \label{tab:adversarial_results_cagn_gat}
  
  \setlength{\tabcolsep}{4pt}
  \renewcommand{\arraystretch}{1}
  
  \sisetup{
    detect-weight=true,
    detect-family=true,
    table-number-alignment = center,
    round-mode=places,
    round-precision=2,
  }
  \begin{tabular}{
    @{}
    l   
    l   
    S   
    S   
    S   
    S   
    S   
    @{}
  }
    \toprule
    \textbf{Attack Type} & \textbf{Param}
      & {\textbf{Acc}}
      & {\textbf{Prec}}
      & {\textbf{Recall}}
      & {\textbf{F1}}
      & {\textbf{AUC}} \\
    \midrule
    Clean
      & --     & 0.86 & 0.95 & 0.74 & 0.83 & 0.96 \\
    \midrule
    \multirow{4}{*}{PGD}
      & $\epsilon=0.1$ & 0.84 & 0.94 & 0.72 & 0.82 & 0.94 \\
      & $\epsilon=0.2$ & 0.84 & 0.93 & 0.71 & 0.81 & 0.92 \\
      & $\epsilon=0.5$ & 0.75 & 0.84 & 0.58 & 0.69 & 0.77 \\
      & $\epsilon=0.7$ & 0.69 & 0.74 & 0.54 & 0.62 & 0.66 \\
    \midrule
    \multirow{4}{*}{EdgeRemove}
      & 1\%    & 0.86 & 0.95 & 0.74 & 0.83 & 0.96 \\
      & 5\%    & 0.86 & 0.95 & 0.74 & 0.83 & 0.96 \\
      & 10\%   & 0.86 & 0.95 & 0.74 & 0.83 & 0.96 \\
      & 20\%   & 0.86 & 0.95 & 0.74 & 0.83 & 0.96 \\
    \midrule
    \multirow{4}{*}{NodeInject}
      & 1\%    & 0.85 & 0.93 & 0.73 & 0.82 & 0.95 \\
      & 5\%    & 0.82 & 0.86 & 0.73 & 0.79 & 0.92 \\
      & 10\%   & 0.79 & 0.79 & 0.71 & 0.75 & 0.89 \\
      & 20\%   & 0.79 & 0.79 & 0.71 & 0.75 & 0.89 \\
    \bottomrule
  \end{tabular}
\end{table}

\subsection{Evaluation Methodology and Experimental Setup}
\label{subsec:evaluation_methodology}

After constructing a unified dataset and establishing a standardized graph representation, this component defines the evaluation protocol in REAL-IoT to assess the robustness of selected GNN-based IDS models against distributional drift, as well as the effectiveness of the LLM-based mitigation.

\begin{figure*}[t!]
  \centering
  \includegraphics[width=1\linewidth]{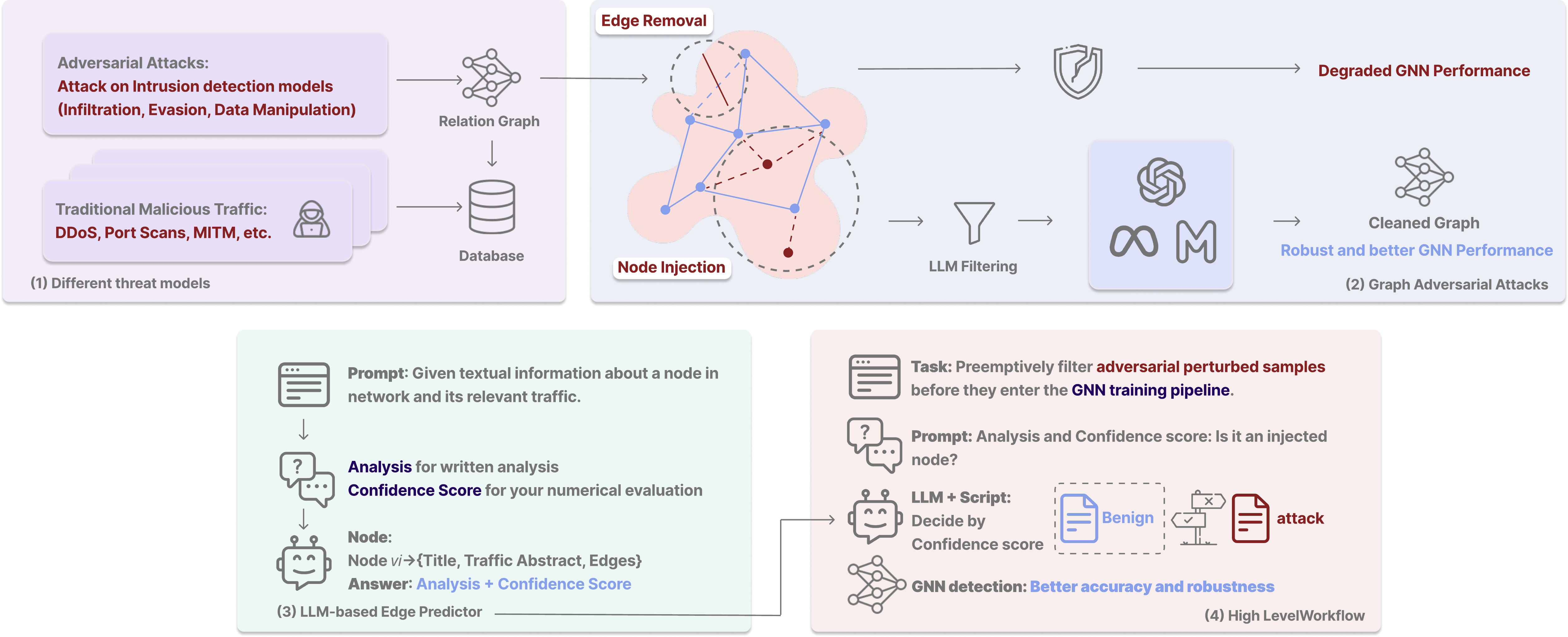}
  \label{fig: LLM}
  \caption{The LLM Mitigation pipeline design}
\end{figure*}

\subsubsection{Evaluation Protocol}
\label{subsubsec:eval_protocol}

\begin{enumerate}
    \item \textbf{Step 1: Baseline Performance Evaluation:} Train and test the GNN model on individual, standardized canonical datasets to establish baseline performance metrics under traditional evaluation conditions.
    \item \textbf{Step 2: Distribution Drift Evaluation:} Utilize the combined standardized dataset. Train the model on specific subsets identified by dataset source and test its performance on distinct subsets. 
    \item \textbf{Step 3: Robustness to Synthetic Adversarial Attacks:}
    The resilience of trained GNN models to adversarial manipulations is assessed using a suite of synthetic attacks. Models (trained on the unified dataset from section ~\ref{subsec:unified_dataset_construction}) are subjected to established white-box attacks. PGD~\cite{Madry18_PGD} is selected as the representative in this category. Then, black-box attack also applied, which includes Edge Removal and Node Injection~\cite{li2020deeprobust}. Attack parameters, such as perturbation budget \( \epsilon \) and the number of iterations, are carefully chosen and specified for each experimental setup. Key metrics are performance degradation under attacks.

    \item \textbf{Step 4: Efficacy of LLM-based Mitigation:}
    The LLM-based mitigation component is evaluated as a preprocessing or analytical layer integrated with the GNN-based NIDS. Its effectiveness is measured by its standalone ability to accurately flag suspicious network flows and, more importantly, by quantifying the performance improvements conferred to the downstream GNN models when operating in collaboration ~\cite{zhang2024can}. This evaluation is performed on relevant splits of the unified dataset to ensure consistency.
\end{enumerate}

\subsubsection{Experimental Setup Details}
\label{subsubsec:experimental_setup_details}

\begin{itemize}
    \item \textbf{Evaluation Tasks:} We primarily focus on binary classification (Attack vs. Benign). For embedding-based GNNs like Anomal-E, representations are learned first, followed by downstream anomaly detection for classification. Multi-class attack categorization is performed as a secondary analysis where applicable.

    \item \textbf{Data Splitting and Preprocessing for Evaluation:} Individual canonical and the unified datasets utilize a stratified 70\%/30\% train/test split. Preprocessing parameters for feature scaling and categorical encoding are derived from the training data to prevent leakage.
\end{itemize}

\subsubsection{Analysis of GNN Robustness to Adversarials}
\label{subsubsec:adversarial_analysis}
The adversarial attack simulations in table ~\ref{tab:adversarial_results_anomal}, ~\ref{tab:adversarial_results_egraphsage}, and ~\ref{tab:adversarial_results_cagn_gat} reveal distinct vulnerabilities and resiliencies across the GNN models. Notably, PGD attacks, representing a white-box approach, induced more significant performance degradation compared to black-box structural attacks. It is particularly evident in the sharp decline in accuracy for Anomal-E and substantial drops in F1-scores for E-GraphSAGE and CAGN-GAT. This underscores the enhanced efficacy of attacks leveraging internal model knowledge. However, the practical applicability of PGD in real-world scenarios is often limited by its white-box prerequisite. All models demonstrated remarkable resilience to edge removal attacks, with performance metrics remaining largely stable even with up to 30\% edge deletion. In contrast, node injection attacks show a more consistent and intuitively scaled degradation across models. This predictable impact suggests that node injection represents a more realistic attack vector for future real-world experimental validation and defensive strategy development.

\subsubsection{LLM-Enhanced GNN Robustness}
\label{subsubsec:llm_robustness_validation}

To assess the novel application of Large Language Models (LLMs) in enhancing GNN robustness against black box adversarial attacks as mentioned in ~\ref{subsubsec:eval_protocol}, we designed a targeted experimental validation. This stage evaluates the LLM's efficacy as an expert system in identifying compromised or victimized original nodes within an adversarially manipulated graph, and the subsequent impact on GNN-based IDS performance.

\textbf{LLM as an Expert Node Analyzer:}
We employ a pre-trained LLM, accessed via various LLM API, to analyze original nodes within a test graph subjected to synthetic node injection. The LLM is prompted with a curated summary for each target node, including its local connectivity profile, distinguishing between edges representing original, standardized flows and edges connected to synthetic attack infrastructure. A particular focus is on the volume and nature of incoming synthetic traffic, a key indicator of the node being targeted.

The LLM's task is to provide a confidence score reflecting the likelihood that the analyzed original node is an attack victim, along with a reasoning. This leverages the LLM's capacity to interpret complex, context-rich descriptions that combine graph-structural cues with flow-level semantics~\cite{zhou2025trustrag}.

\textbf{Mitigation and GNN Performance Impact:}
Original nodes flagged by the LLM are considered identified victims. To simulate mitigation, these nodes and their incident edges are pruned from the attacked graph, creating an "LLM-fixed" graph. The core of this experimental stage is to then evaluate the GNN's intrusion detection performance (using metrics detailed in Section~\ref{subsubsec:experimental_setup_details}) on three distinct graph conditions:
\begin{enumerate}
    \item The original, clean test graph (baseline).
    \item The test graph with injected attacks (impact).
    \item The LLM-fixed test graph (mitigation effectiveness).
\end{enumerate}
This comparative analysis directly quantifies the LLM's contribution to restoring GNN performance in adversarial settings by accurately identifying and enabling the neutralization of compromised or heavily targeted network entities. This approach uniquely combines graph-based anomaly detection with LLM-driven expert analysis for enhanced IDS resilience.

\begin{table}[t]
\centering
\caption{Performance of LLMs in graph mitigation with 100 sample nodes and 20 injected nodes. Node count in graph after LLM processing, CF = Correctly Flagged, IF = Incorrectly Flagged, Downstream GNN model: CAGN-GAT Fusion}
\label{tab:merged_results}

\renewcommand{\arraystretch}{1.2}
\setlength{\tabcolsep}{4pt}
\sisetup{
  detect-weight        = true,
  detect-family        = true,
  round-mode           = places,
  round-precision      = 3,
  table-number-alignment = center
}
\begin{adjustbox}{max width=\columnwidth}
\begin{tabular}{
  @{} l
  S[table-format=1.3]   
  S[table-format=1.3]   
  S[table-format=1.3]   
  S[round-precision=0, table-format=3.0]  
  S[round-precision=0, table-format=2.0]  
  S[round-precision=0, table-format=2.0]  
  @{}
}
\toprule
\textbf{Model} 
  & {\textbf{Accuracy}} 
  & {\textbf{F1}} 
  & {\textbf{LLM Recall}} 
  & {\textbf{Nodes}} 
  & {\textbf{CF}} 
  & {\textbf{IF}} \\
\midrule
Baseline             
  & 0.650 & 0.720 & {--} & 100 & {--} & {--} \\
Claude 3.5 Haiku~\cite{anthropic2024claude35haiku}   
  & 0.653 & 0.661 & 0.851 & 118 & 0.698 & 2 \\
DeepSeek R1~\cite{deepseek2025r1}         
  & 0.608 & 0.608 & \bfseries 0.960 & 79 & 1 & 40 \\
Claude 3.7 Sonnet~\cite{anthropic2025claude37sonnet}   
  & 0.627 & 0.655 & 0.857 & 102 & 4 & 14 \\
LLaMA 4 Maverick 17B~\cite{meta2025llama4maverick}             
  & 0.769 & \bfseries 0.815 & 0.917 & 65 & \bfseries 20 & 35 \\
GPT-4o~\cite{openai2024gpt4o}            
  & \bfseries 0.820 & 0.770 & 0.846 & 86 & \bfseries 20 & \bfseries 14 \\
\bottomrule
\end{tabular}
\end{adjustbox}
\end{table}

\subsubsection{Analysis of LLM-based Graph Mitigation Performance}
Our evaluation of LLMs for graph mitigation, tested at 100-node (20 injected) and 1000-node (200 injected) scales with CAGN-GAT as the downstream NIDS model, revealed that LLaMA 4 Maverick 17B and GPT-4o consistently enhanced downstream GNN performance. For instance, LLaMA 4 achieved a peak F1-score of 0.815 on the smaller graph (baseline 0.720) and GPT-4o reached 0.838 on the larger graph (baseline 0.821), effectively identifying malicious nodes. However, this uplift was less significant on the larger graph, indicating diminishing returns. Meanwhile, some LLMs presented challenges: DeepSeek R1, despite a high reported LLM recall (0.960), ultimately degraded GNN performance (F1 0.608) due to very low precision (1 Correctly Flagged vs. 40 Incorrectly Flagged), illustrating that high recall alone can be misleading. Furthermore, models like Claude 3.5 Haiku struggled significantly, especially on the larger dataset where it failed to identify any malicious nodes (CF=0) while incorrectly flagging 164 benign ones (resulting in F1 0.673), underscoring the variability in LLM efficacy and the potential for performance degradation.

\subsubsection{Limitations of LLM Evaluation}
The LLM evaluation was constrained by several factors: experiments were conducted on sampled subgraphs due to significant API costs (e.g., \$0.05 per call for GPT-4o) and time efficiency issues (e.g., DeepSeek R1 averaging 60 seconds per prompt), primarily using the CAGN-GAT model for this initial test of LLM mitigation, which may limit generalizability to other GNNs or full-scale datasets. We observed diminishing performance returns from LLM mitigation at a larger scale. Furthermore, results are inherently tied to our specific prompt engineering, and the interpretation of LLM efficacy metrics in relation to final GNN performance is complex. The dynamic nature of LLMs also means these findings might vary over time.

\section{Controlled Physical IoT Testbed}

\subsection{Motivation and Objectives}
\label{subsec:testbed_motivation}
After the measurements conducted in the previous section where we used REAL-IoT to evaluate both individual and unified canonical datasets, we discovered that existing public datasets suffer from lacking realism in representing contemporary IoT device behaviors. For surveyed GNN-based NIDS, they also have shortcomings such as being outdated or not being developed and benchmarked with practical adversarial scenarios in mind. Also, we want to validate Real-IoT under a more realistic setting. To address this issue, we found it important to generate a new dataset within a controlled environment, providing a known ground truth for specific attack scenarios. In section~\ref{subsubsec:adversarial_analysis} it also shows that as a ramification of IP spoofing and random-sourced DoS, Node Injection is a pertinent and intuitive attack vector in a real-world context. Therefore, DoS attacks are featured in this experiment design.

\begin{table}[t]
\centering
\caption{Performance of LLMs in graph mitigation with 1000 sample nodes and 200 injected nodes. Node count in graph after LLM processing, CF = Correctly Flagged, IF = Incorrectly Flagged Downstream GNN model: CAGN-GAT Fusion}
\label{tab:new_results}

\renewcommand{\arraystretch}{1.2}
\setlength{\tabcolsep}{4pt}
\sisetup{
  detect-weight        = true,
  detect-family        = true,
  round-mode           = places,
  round-precision      = 3,
  table-number-alignment = center
}
\begin{adjustbox}{max width=\columnwidth}
\begin{tabular}{
  @{} l
  S[table-format=1.3]   
  S[table-format=1.3]   
  S[table-format=1.3]   
  S[round-precision=0, table-format=4.0]  
  S[round-precision=0, table-format=3.0]  
  S[round-precision=0, table-format=3.0]  
  @{}
}
\toprule
\textbf{Model} 
  & {\textbf{Accuracy}} 
  & {\textbf{F1}} 
  & {\textbf{LLM Recall}} 
  & {\textbf{Nodes}} 
  & {\textbf{CF}} 
  & {\textbf{IF}} \\
\midrule
Baseline               
  & 0.842 & 0.821 & {--} & 1000 & {--} & {--} \\
Claude 3.5 Haiku~\cite{anthropic2024claude35haiku}   
  & 0.777 & 0.673 & 0.698 & 1036 & 0 & 164 \\
Claude 3.7 Sonnet~\cite{anthropic2025claude37sonnet}   
  & 0.774 & 0.728 & 0.741 & 1107 & 35 & \bfseries 58 \\
LLaMA 4 Maverick 17B\cite{meta2025llama4maverick}              
  & \bfseries 0.859 & 0.834 & \bfseries 0.758 & 931 & \bfseries 200 & 69 \\
GPT-4o~\cite{openai2024gpt4o}                
  & 0.857 & \bfseries 0.838 & 0.774 & 945 & 197 & \bfseries 58 \\
\bottomrule
\end{tabular}
\end{adjustbox}
\end{table}
The primary objectives of building this testbed dataset are:
\begin{enumerate}
    \item To design a small-scale yet representative physical testbed emulating an IoT network, including simulated IoT devices and a dedicated attack platform.
    \item To execute a series of carefully orchestrated and reproducible network attacks, focusing on DoS variants and IP spoofing, against the simulated IoT devices.
    \item To capture the resulting network traffic, process it into bi-directional NetFlow records, and perform meticulous labeling based on ground truth to create a new, albeit modest, validation dataset.
    \item To leverage this dataset to validate the performance of selected GNN models which previously evaluated on public datasets, thereby assessing their efficacy under specific, controlled attack conditions.
\end{enumerate}

\begin{figure}[t!] 
  \centering
  \includegraphics[width=\linewidth]{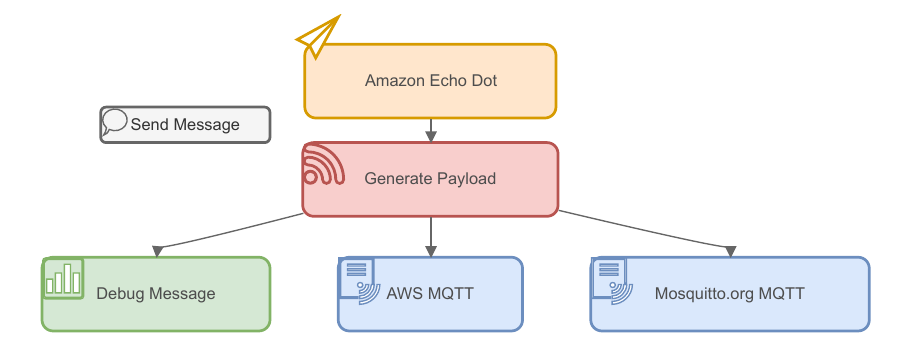} 
  \caption{Node-Red flow of Echo Dot Traffic}
  \label{fig:NodeRed} 
\end{figure}

\subsection{Testbed Setup and Architecture}
\label{subsec:testbed_architecture}
\begin{figure}[t!] 
  \centering
  \includegraphics[width=\linewidth]{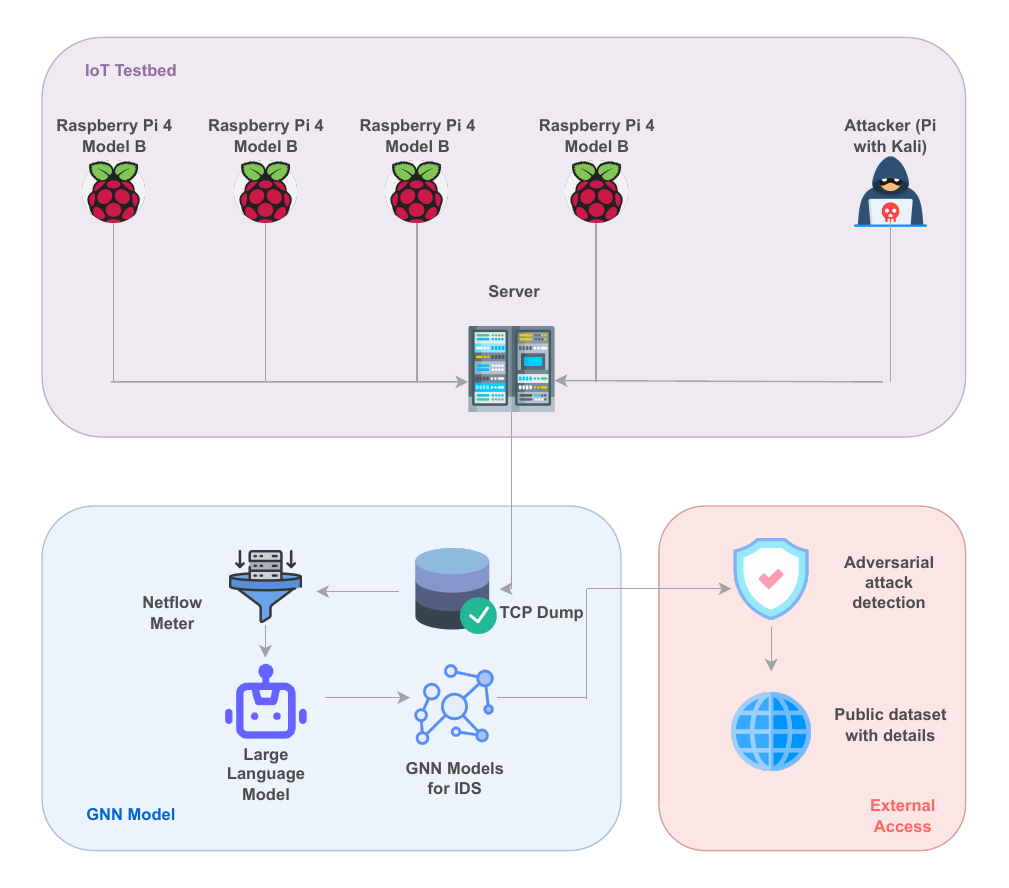} 
  \caption{High level lab setup diagram}
  \label{fig:lab} 
\end{figure}

Our physical testbed was constructed using readily available hardware to simulate a typical small-scale IoT Smart Home environment. The setup, depicted in Figure~\ref{fig:lab}, consists of five Raspberry Pi devices. One Raspberry Pi 4B was configured with Kali 2024.4 ~\cite{kali2024_4} to serve as the dedicated attacker node, with the remaining four Raspberry Pi 4B devices tasked with simulating IoT device behavior. 

In the original design, an Amazon Echo Dot 2nd Generation was connected to a Raspberry Pi for realistic MQTT communication. However, due to the limited support of Amazon Alexa for Raspberry Pi OS~\cite{raspberrypios64}, and the restraints in Raspberry Pi 4B's computing performance~\cite{kolcun2020case}, we decide to capture the (encrypted) net traffic from the Echo Dot and use Node-Red to mimic the same behavior and payload directly on the Pi to a designated, pre-registered AWS MQTT broker on AWS IoT Greengrass~\cite{aws2025greengrass}, which ensures best performance while keeping the realism of the physical experiments.

\begin{table}[htbp] 
\centering
\caption{Configuration of Devices in the Physical Testbed.}
\label{tab:device_configurations}
\resizebox{\columnwidth}{!}{
\begin{tabular}{llll}
\toprule
Role & Device Model & Operating System & IP Address \\
\midrule
Attacker & Raspberry Pi 4B & Kali Linux 2024.4 & 192.168.2.4 \\
IoT Sim 1 (PI\_OS\_1) & Raspberry Pi 4B & Raspberry Pi OS (64-bit) & 192.168.2.3 \\
IoT Sim 2 (PI\_OS\_2) & Raspberry Pi 4B & Raspberry Pi OS (64-bit) & 192.168.2.6 \\
IoT Sim 3 (PI\_OS\_3) & Raspberry Pi 4B & Raspberry Pi OS (64-bit) & 192.168.2.8 \\
IoT Sim 4 (PI\_OS\_4) & Raspberry Pi 4B & Raspberry Pi OS (64-bit) & 192.168.2.15 \\
\bottomrule
\end{tabular}%
}
\end{table}

Network connectivity for the testbed was provided by a MacBook Pro 14-inch, 2024 running macOS Sequoia 15.4.1. The Mac was configured as a Wi-Fi access point (AP) using macOS's built-in Internet Sharing feature, creating an isolated wireless network. The Mac also functioned as the gateway and DHCP server for this internal network.

Each of the four Raspberry Pi OS devices employed Node-RED 4.0.9~\cite{nodered409} to simulate IoT client activities. These clients were programmed to periodically generate network traffic by attempting to send MQTT messages on port 8883 ~\cite{nodered}. This setup aimed to generate a baseline of benign communication attempts at the application layer, mimicking typical interactions between IoT devices and cloud services. The Kali Linux attacker was equipped with standard network utilities, primarily hping3 3.0.0~\cite{hping3} for DoS traffic generation and nmap 7.95~\cite{nmap795} for network scanning.

\subsection{Data Collection and Attack Scenarios}
\label{subsec:testbed_datacollection}
Network traffic within the testbed was captured on the Mac AP. We utilized Wireshark 4.4.6 ~\cite{wireshark446} to capture all packets on the \texttt{bridge100} interface, where the subnet constructed by the Mac is located. The raw data was saved in PCAP format.

The data collection was structured into distinct 5-minute phases, orchestrated by a Bash script executed on the Kali attacker. A detailed, timestamped log of all actions was maintained by this script. The sequence of phases was as follows:

\begin{enumerate}
    \item \textbf{Benign Phase (Start, 0--5 minutes):} All four Node-RED Pi devices generated HTTP traffic and simulated benign MQTT traffic. No attack activity was present during this initial phase.

    \item \textbf{Attack B -- IP Spoofing DoS (5--10 minutes):} This phase targeted PI\_OS\_2 (192.168.2.6). The attack utilized \texttt{hping3} with the \texttt{--rand-source} option to send a high volume of TCP SYN packets to the SSH port. Based on preliminary tests, the source ports for these attack packets were observed to start from a base port of 60002 and increment with successive packets or differing spoofed source IP addresses.

    \item \textbf{Attack C -- Nmap Port Scan (10--15 minutes):} An \texttt{nmap} TCP SYN scan was launched from the Kali attacker against a specific range of TCP ports on PI\_OS\_3. An attempt was made to mark scanner traffic using \texttt{nmap}'s \texttt{-g 60003} option.

    \item \textbf{Attack D -- Combined DoS:} This multi-vector DoS attack targeted PI\_OS\_4 (192.168.2.15). It involved a simultaneous flood of TCP SYN, UDP, and ICMP packets. For the TCP and UDP components, source ports were configured to initiate from a base of 60004 and increment. Target ports for these TCP and UDP streams were also set to increment.

    \item \textbf{Benign Phase (End, 20--25 minutes):} In the final phase, all attack activities were ceased, and the network returned to a state where only HTTP and benign Node-RED MQTT simulation traffic was present.
\end{enumerate}

Network traffic was processed into bidirectional NetFlow records using CICFlowMeter (v0.3.0)~\cite{cicflowmeter}. The final 4161 record NetFlow dataset was assembled from labeled segments of all three sessions detailed in Table~\ref{tab:experiment_chronology_final} in the Appendices. Labeling relied on timestamped logs and attack parameters: direct Kali IP for Nmap, and target IP/port combined with observed incrementing source port patterns for DoS attacks. All other flows were labeled benign.

\subsection{Feature Engineering for Compatibility}
\label{subsec:testbed_feature_eng}
To ensure the NetFlow data generated from our testbed was compatible with selected GNN models, we perform feature engineering. The task is to map the features output by CICFlowMeter to the schema expected by these models, particularly aligning with the NF-V2-series feature set.

A key feature, L7\_PROTO (Application Layer Protocol), is not directly generated by CICFlowMeter. We engineered this feature by mapping well-known destination or source port numbers to their corresponding application-layer protocol identifiers. This mapping was guided by the nDPI library's protocol ID list \cite{nDPI_GH_Link_or_Ref}, a common reference for deep packet inspection based protocol identification. Flows associated with ports not in our mapping list were assigned an 'Unknown' L7 protocol category (converted to numerical value 0).

It is important to note that for the data intake of Anomal-E model, which requires a specific set of 40 features often generated by proprietary tools like nProbe ~\cite{deri2003nprobe}, which are reflected in the feature set and characteristics of NF-V2 series of data like NF-UNSW-NB15-v2~\cite{Sarhan22_StandardFeatureSet}. A complete feature mapping from our open-source CICFlowMeter output was not achieved due to the unavailability of granular features and the closed-source nature of nProbe. Consequently, results for Anomal-E on this specific testbed dataset are not presented in this work to align with our commitment to utilizing reproducible open-source methodologies where possible.

\begin{table}[htbp]
\centering
\caption{GNN Model Performance on the Testbed Dataset.}
\label{tab:testbed_gnn_performance}
\resizebox{\columnwidth}{!}{%
\begin{tabular}{lcccccc}
\toprule
Model & AUC & Accuracy & F1-Score & Precision & Recall  \\ 
\midrule
EGraphSage & 0.9113 & 0.9370 & 0.9135 & 0.8688 & 0.9631 \\
CAGN-GAT Fusion & 0.8567 & 0.8827 & 0.8502 & 0.7606 & 0.9638 \\
\bottomrule
\end{tabular}%
}
\end{table}

\subsection{Validation Results and Analysis}
\label{subsec:testbed_results}
The performance of E-GraphSAGE (F1-Score: 0.9135) and CAGN-GAT Fusion (F1-Score: 0.8502) on our previously unseen testbed dataset (Table~\ref{tab:testbed_gnn_performance}) demonstrates a good level of generalization, despite these models not being trained on this specific data. While these scores do not reach the state-of-the-art levels reported on their original public benchmark datasets (as in Table~\ref{tab:gnn_benchmark}), the strong recall (0.9631 for E-GraphSAGE, 0.9638 for CAGN-GAT) and respectable overall F1-scores are significant. This suggests that the labeling of our testbed data is largely coherent and captures meaningful distinctions between benign and attack traffic, enabling the models to identify a high proportion of threats. The moderate performance degradation compared to original benchmarks points to a realistic and manageable degree of distribution drift. This scenario confirms the testbed's utility as a challenging yet valid environment for evaluating GNN-based NIDS.

\subsection{Limitations of Testbed and Future work}
\label{subsec:testbed_summary_limitations}

The testbed evaluation has several limitations. The generated dataset is modest in scale (4161 NetFlow records) and represents a short observation period (approximately 60 minutes of effective traffic). The IoT device simulation was limited to Node-RED based MQTT communication attempts and does not capture the full diversity of real-world IoT application-layer interactions or device heterogeneity. The attacks, though designed to emulate common tactics, consist of only a small subset of the broader threat landscape. 

Despite these limitations, this physical testbed evaluation offers valuable preliminary insights and serves as a methodological contribution towards creating reproducible, labeled datasets for NIDS research. It complements evaluations on larger public datasets by providing a focused perspective on model performance against specific, verifiable attacks. 

Future work will aim to expand the scale and diversity of both benign and attack traffic in the testbed, incorporate more sophisticated attack vectors, refine the data processing pipeline, and improve the LLM mitigation design. We believe this effort will help the development of
more resilient and trustworthy GNN-based IDS in practice.

\section{Conclusion}
In this paper, we identified limitations in evaluating GNN-based IDS for IoT environments: the prevalent use of single, preprocessed datasets that overlook real-world distribution drift and lead to models overfitting, and reliance on synthetic adversarial attacks that lack realism. To address these issues, we introduced REAL-IoT, a comprehensive framework for more realistic and holistic GNN-based NIDS robustness assessment. REAL-IoT achieves this through its key components: a methodology for creating a Unified Dataset constructed from canonical NIDS datasets to analyze model generalization and drift susceptibility; a novel, IoT-focused intrusion dataset collected from our physical testbed featuring practical adversarial attacks; a multi-stage evaluation protocol assessing performance under varied conditions; and an exploration of LLM-based strategies for attack mitigation. 

Our findings reveal performance discrepancies often missed by existing evaluation approaches and emphasize the need to consider both distribution drift and practical adversarial attacks. The REAL-IoT framework hence represents a step towards more robust evaluation practices, aiming to advance the development of resilient GNN-based NIDS for IoT.

\section{Artifact Availability}

To support reproducibility, we provide anonymized code and data at
\href{https://anonymous.4open.science/r/REAL-IoT-0F69/}{this anonymous GitHub repository}.
All artifacts will be made publicly available upon publication.

\bibliographystyle{ACM-Reference-Format}
\bibliography{base}

\clearpage
\appendix

\section{Ethics}
This work does not raise any ethical concerns. All experiments were conducted using simulated network traffic generated within a controlled testbed environment. The network was fully isolated and operated in a private local setting, with no connection to or data exposure on the public Internet. No human subjects or real user data were involved at any stage of the study.

\section{Physical Setup of the IoT testbed}
\begin{figure}[t!] 
  \centering
  \includegraphics[width=\linewidth]{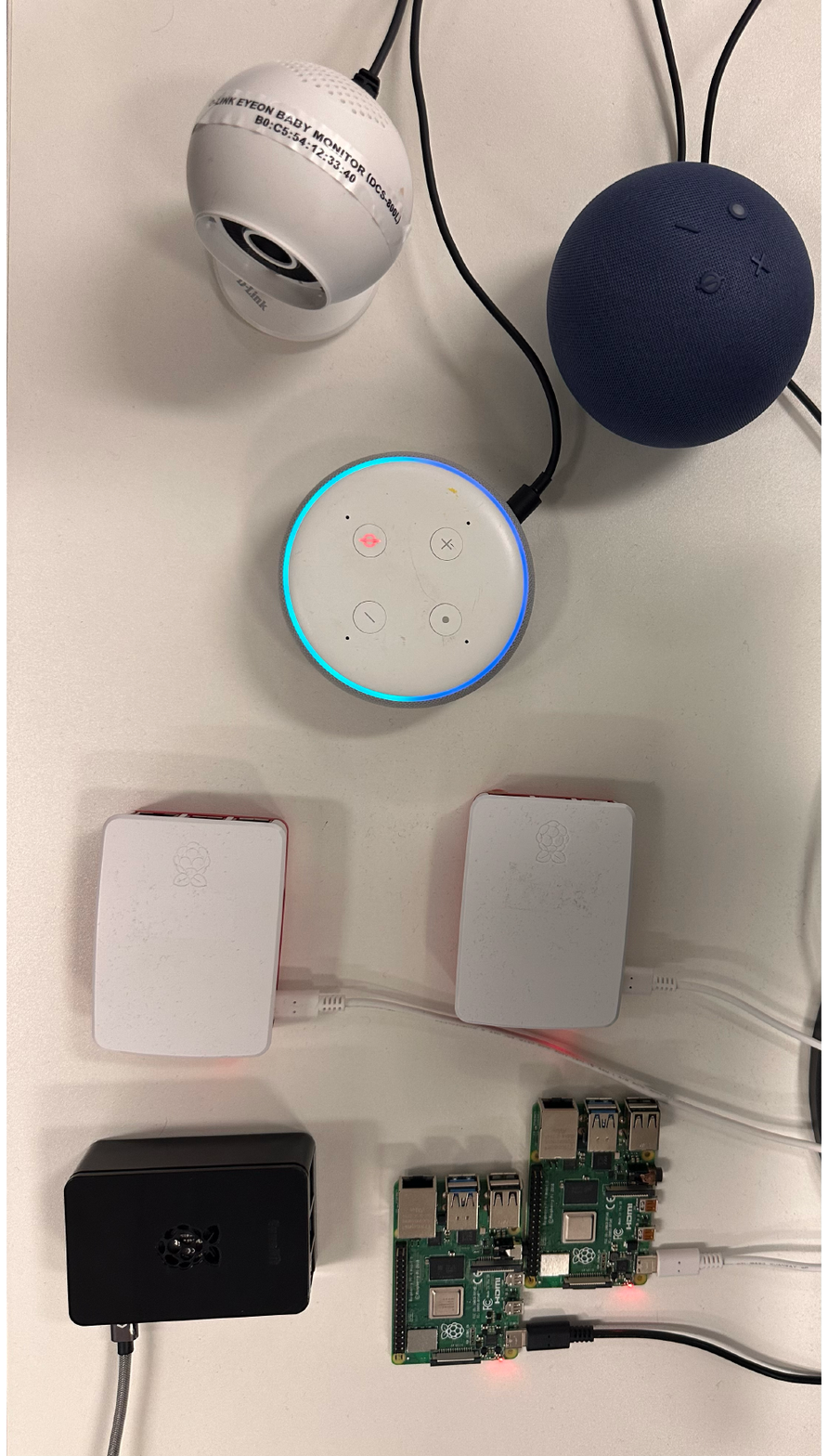} 
  \caption{Physical Testbed setup}
  \label{fig:Testbed} 
\end{figure}

Figure~\ref{fig:Testbed} shows a photograph of the actual testbed setup. Five Raspberry Pi 4B devices are visible, alongside an Amazon Echo Dot (2nd Generation). 

The figure captures a moment during which Wireshark is used to monitor packets transmitted by the Echo Dot, allowing us to replicate and analyze its network behavior. 

\section{Common NetFlow Feature Set}
Table~\ref{tab:netflow_features_adjusted} presents the standardized common NetFlow feature set adopted across the NF-V2 series datasets within the REAL-IoT framework. This feature schema was created to serve as a unified representation layer, unifying data collected from heterogeneous sources and ensuring consistency across benchmark experiments. The goal of this standardization is to support consistent graph construction, model input formatting, and comparative evaluations across datasets, models, and attack scenarios.Beyond raw traffic metadata, the schema includes auxiliary fields used for experiment control and supervision. The \texttt{flow\_id} acts as a unique identifier to support traceability, data identity, and reproducibility. The \texttt{dataset\_source} tag is used to distinguish between flows originating from different datasets or collection environments within the REAL-IoT framework.

\begin{table}[t]
\centering
\caption{Standardized Common NetFlow Feature Set}
\label{tab:netflow_features_adjusted} 

\renewcommand{\arraystretch}{1.3} 
\setlength{\tabcolsep}{4pt}
\begin{adjustbox}{max width=\columnwidth}

\begin{tabular}{@{} l p{4cm} @{}} 
\toprule
\textbf{Feature Name} & \textbf{Description} \\
\midrule
\texttt{IPV4\_SRC\_ADDR} & Source IPv4 Address. \\
\texttt{L4\_SRC\_PORT} & Layer 4 Source Port. \\
\texttt{IPV4\_DST\_ADDR} & Destination IPv4 Address. \\
\texttt{L4\_DST\_PORT} & Layer 4 Destination Port. \\
\texttt{PROTOCOL} & IP Protocol Number. \\
\texttt{L7\_PROTO} & Layer 7 Protocol. \\
\texttt{IN\_BYTES} & Total incoming bytes. \\
\texttt{OUT\_BYTES} & Total outgoing bytes. \\
\texttt{IN\_PKTS} & Total incoming packets. \\
\texttt{OUT\_PKTS} & Total outgoing packets. \\
\texttt{TCP\_FLAGS} & TCP connection flags. \\
\texttt{FLOW\_DURATION\_MILLISECONDS} & Flow duration. \\
\midrule 
\texttt{flow\_id} & Unique identifier for flow. \\
\texttt{dataset\_source} & Identifier for the dataset. \\
\texttt{Attack} & Binary: Attack or Benign. \\
\texttt{Label} & Specific attack category. \\
\bottomrule
\end{tabular}
\end{adjustbox}
\end{table}

\section{LLM Mitigation Prompt}
This prompt is designed to guide a LLM in analyzing raw NetFlow traffic represented as a graph. The LLM is tasked with evaluating whether a given node appears anomalous or adversarial based on raw feature consistency within its local neighborhood. The analysis enables the model to identify potentially malicious behavior such as evasion attacks or adversarial perturbations. This is a core design of the LLM-Enhanced GNN Robustness component of REAL-IoT as described in ~\ref{subsubsec:llm_robustness_validation}

\begin{figure*}[t]
\centering
\begin{tcolorbox}[top=2pt, bottom=2pt, left=4pt, right=4pt]
\textbf{System prompt:} You are an expert cybersecurity analyst reviewing network flow data represented as a graph. The graph contains {num\_total\_nodes} nodes (flows) representing network activity. Some nodes may represent normal activity, while others could be anomalous or potentially malicious (e.g., part of DDoS, spoofing, reconnaissance.Your task is to analyze the single node detailed below. Focus on comparing the node's raw features (if available) to its neighbors' raw features to assess consistency within its local neighborhood. Also consider if the node's own raw features (or processed features if raw are unavailable) seem inherently unusual.Provide a confidence score (0.0 = consistent/likely normal, 1.0 = inconsistent/highly likely anomalous) and a brief justification.\\
Key Analysis Points:\\
Raw Feature Consistency: Does the target node's raw feature values align with its neighbors' raw values? (Primary focus)\\
Unusual Own Features: Does the target node have strange raw feature combinations (e.g., high packets for low-traffic protocol, impossible flags)?\\
Connectivity: Does the node connect to neighbors that seem unusually diverse or disconnected? (Secondary consideration)\\
Key Raw Features (Used for Comparison):\\
'IN\_BYTES': "Incoming bytes", 'OUT\_BYTES': "Outgoing bytes",
        'IN\_PKTS': "Incoming packets", \\
        'OUT\_PKTS': "Outgoing packets",\\
        'FLOW\_DURATION\_MILLISECONDS': "Duration of the flow (ms)",\\
        'PROTOCOL': "Network protocol (e.g., 6=TCP, 17=UDP)",\\
        'L7\_PROTO': "Application layer protocol (numeric code)",\\
        'TCP\_FLAGS': "TCP flags set (numeric code)"\\
\textbf{User content:} Node $v_i$$\rightarrow$\{Title, Raw Traffic, Related Traffic (Edges)\}.
\end{tcolorbox}
\end{figure*}

\begin{table*}[!htbp] 
    \centering

    \caption{Data Collection Experiment Chronology and Phases} 
    \label{tab:experiment_chronology_final}
    \rowcolors{2}{gray!10}{white} 
    \begin{tabularx}{\textwidth}{ 
        >{\raggedright\arraybackslash}p{1.2cm} 
        >{\raggedright\arraybackslash}p{2.2cm} 
        >{\centering\arraybackslash}p{1.6cm} 
        >{\centering\arraybackslash}p{1.6cm} 
        >{\centering\arraybackslash}p{1.5cm} 
        >{\raggedright\arraybackslash}p{3.5cm} 
        >{\raggedright\arraybackslash}X        
    }
    \toprule
    \rowcolor{white} 
    \textbf{Date} & \textbf{Session Phase} & \textbf{Start (UTC)} & \textbf{End (UTC)} & \textbf{Duration} & \textbf{Activity / Attack Type} & \textbf{Target Pi \& Key Params/Notes} \\
    \midrule
    Day 1 & Benign (S1-P0) & 13:46:00 & 13:56:00 & 10 min & Benign Traffic & All Pis normal operation \\
     & Attack (S1-P1) & 13:56:00 & 14:16:00 & 20 min & IP Spoofing DoS & PI\_OS\_2; `--rand-source`, `-s 60002` (start port) \\
    \midrule 
    Day 2 & Benign (S2-P0) & 20:50:01 & 20:55:01 & 5 min & Benign Traffic & Baseline \\
     & Attack (S2-P1) & 20:55:01 & 21:00:01 & 5 min & IP Spoofing DoS & PI\_OS\_2; `--rand-source`, `-s 60002` (start port) \\ 
     & Attack (S2-P2) & 21:00:01 & 21:05:01 & 5 min & Nmap Port Scan & PI\_OS\_3; `-sS -T4 -p [...] -g 60003` \\
     & Attack (S2-P3) & 21:05:01 & 21:09:05 & $\sim$4 min & Combined DoS & PI\_OS\_4; `--rand-source`, `-s 60004` (start), TCP/UDP/ICMP \\
    \midrule 
    Day 3 & Benign (S3-P0) & 22:07:00 & 22:12:00 & 5 min & Benign Traffic & Baseline \\
     & Attack (S3-P1) & 22:12:00 & 22:17:00 & 5 min & UDP DoS & [Target Pi for UDP]; `--rand-source`, `-s 60005` (start) \\
     & Attack (S3-P2) & 22:17:00 & 22:21:35 & $\sim$5 min & ICMP DoS & [Target Pi for ICMP]; `--rand-source` \\
    \bottomrule
    \end{tabularx}
\end{table*}

\end{document}